\documentclass[prc,nofootinbib,twocolumn]{revtex4}
\usepackage{graphicx,dcolumn,array,bm,mathrsfs}

\newcommand{\svec}[1]{\bm{#1}}

\begin{document}
\bibliographystyle{apsrev}


\title{Effective Operators for Double-Beta Decay}
\author{J. Engel}
\email[]{engelj@physics.unc.edu}
\affiliation{Department of Physics and Astronomy, CB3255,
             University of North Carolina, Chapel Hill, North Carolina
             27599-3255}
\author{P. Vogel}
\email[]{vogel@citnp10.caltech.edu}
\affiliation{Kellogg Radiation Laboratory and Physics Department,
             Caltech, Pasadena, CA  91125}
\date{\today}
\begin{abstract}
We use a solvable model to examine double-beta decay, focusing on the 
neutrinoless mode.  After examining the ways in which the neutrino propagator 
affects the corresponding matrix element, we address the problem of finite 
model-space size in shell-model calculations by projecting our exact wave 
functions onto a smaller subspace.  We then test both traditional and more 
recent prescriptions for constructing effective operators in small 
model spaces, 
concluding that the usual treatment of double-beta-decay operators in 
realistic calculations
is unable to fully account for the neglected parts of the model space.  We also
test the quality of the Quasiparticle Random Phase Approximation and 
examine a recent proposal within that framework to use two-neutrino decay to fix
parameters in the Hamiltonian.   
The procedure eliminates the dependence of neutrinoless decay on some unfixed parameters and reduces
the dependence on model-space size, though it doesn't eliminate the latter
completely.  
\end{abstract}
\maketitle
%
%
\section{Introduction}
\label{s:intro}

Next-generation double-beta-decay experiments will provide data on the mass
of the lightest neutrino \cite{elliot02}.  But to interpret the data
accurately we will need to better understand the nuclear physics that
influences the decay rate.  The problem of calculating double-beta-decay
matrix elements has occupied nuclear theorists for a long time.  The latest
in a long history of neutrinoless-decay calculations are thought to approximate 
exact results
to within factors of 3 or 4, but that estimate is not much more than an
informed guess.  In short, the calculations need to be improved.

Almost all current calculations are performed either in the Quasiparticle
Random Phase Approximation (QRPA) or the shell model.  The two methods
differ in the types of correlations they take into account.  The QRPA can
include the cumulative effects of small numbers of particles or holes far
above or below the Fermi surface, but the correlations are of a specific
type.  The shell model is restricted to configurations with low excitation,
but correlates them in all possible ways.  The QRPA calculations indicate
that nearly empty orbits relatively far from the Fermi surface play a 
significant role in the nuclear ground states, implying that the shell-model is 
leaving something out.  
We are not referring here to the short-range correlations coming from 
the nucleon hard core, the representation of which requires very high-lying 
excitations.  
QRPA correlations are longer range, connected to orbits just 10 or 
20 MeV away from the Fermi surface; the shell model usually cannot include even 
these.  Within the levels it does include, however, the shell model indicates 
that QRPA correlations are too simple.  We need either to add the effects of 
more complicated correlations to the QRPA or to include effects of additional
single-particle levels in the shell model (or perhaps do both).

The way to correct for the finite shell-model space is to construct
effective Hamiltonians and transition operators.   Unfortunately, the
equations that determine effective operators are hard to solve and
perturbative approximations often do not work.  Phenomenology is the usual
substitute.  Thus, the effective Hamiltonians computed within perturbation
theory usually include only two-body terms that are then renormalized to
fit ground-state energies and other observables.  Effective transition
operators are often taken to have the same form as the bare operators but
are multiplied by effective coupling constants. Such prescriptions can work
well when there are lots of data to which parameters can be fit, but for
the two-body operator governing neutrinoless double-beta decay, which
has not yet been observed at all, the use of the bare operator with or
without simple multiplicative renormalization is suspect.  Yet such is the
state of the art in shell-model calculations of double-beta decay. To
improve these calculations, we should focus on a better determination of
the effective decay operator.

Here we use a solvable model based on the algebra $SO(5)$ to test some of
the assumptions that go into calculations of double-beta decay
in $^{76}$Ge, one of the most promising isotopes for experiment.  We diagonalize
the model Hamiltonian and calculate the rates of both two-neutrino ($2\nu$) and
neutrinoless ($0\nu$) decay in a space consisting of a degenerate $pfg_{9/2}$
valence shell, and a higher-energy degenerate $sdg_{7/2}$ shell.  In this
quite
realistic model space (albeit with unrealistically simple correlations), we
can compare the behavior of the two kinds of decays, which depend
differently on radial wave functions, as we vary couplings in the
Hamiltonian.  We can also project the full wave function onto states in which
all nucleons are in the lower
shell and test procedures for constructing effective interactions and
operators. Though technically we are calculating Fermi decay, we allow
isospin to be violated so that our calculation mimics many of the aspects
of Gamow-Teller decay. We find, for example, that an effective two-body
operator that reproduces decay rates exactly in nuclei with just a few
valence nucleons, while better than the bare
operator, is far from perfect when applied to a nucleus like $^{76}$Ge that has 
many valence
particles.  The more phenomenological prescription of multiplying the bare
operator by an effective coupling requires different renormalization
factors for $2\nu$ and $0\nu$ decay. Such results, while
perhaps unsurprising, contradict common practice.

We also examine some related ideas.  A recently proposed technique for
constructing effective operators nonperturbatively through a Lanczos
approximation to the many-body Green's function usually proves efficient but
appears to offer little advantage over the Lanczos methods
already used in shell-model calculations for double-beta decay.  A recent
proposal for adjusting the Hamiltonian in the QRPA to fit measured $2\nu$
decay,
though it eliminates the dependence on some of the parameters in the
Hamiltonian and reduces the dependence on model-space size,
does not work as well in this model as in more realistic calculations. 
The QRPA itself, ironically, reproduces exact solutions extremely 
accurately here.

The rest of this paper is divided as follows:  In section
\ref{s:2} we discuss the model.  Though it has been used before in
$2\nu$ decay \cite{civitarese97}, ours is the first application to $0\nu$ decay,
the operator for which is not built from SO(5) generators and therefore
takes some effort to manipulate.  Section \ref{s:3} tests common
prescriptions for constructing effective operators in a truncated model
space against exact solutions in the full model space.  In section
\ref{s:4} we examine the proposal of Ref.\ \cite{rodin03} to renormalize the
Hamiltonian in the QRPA.  Section \ref{s:5} is a conclusion.  The appendix
contains explicit formulae for matrix elements of double-beta decay
operators in the model.
%
\section{SO(5) model}
\label{s:2}

We consider the nuclei $^{76}$Ge and $^{76}$Se in a ``small" shell-model space
containing 12 protons and 24 neutrons (for Ge) in a
degenerate $pfg_{9/2}$ valence shell, and in
a ``large" space containing the small space together with all possible
excitations of the nucleons into a degenerate version of
the next shell ($sdg_{7/2}$), which is an energy $\epsilon$ above the
$pfg_{9/2}$ shell.  We 
usually use one of two different values of $\epsilon$ so that either about 20\%
of the 
particles are in the upper levels, or about 6\% .  Both values of the splitting, 
probably, are unrealistically small for the kind of long-range correlations we 
have in mind, but our correlations are unrealistically simple and we compensate 
in part by lowering $\epsilon$ (for one test we will use a very large $\epsilon$
that
results in 1\% of the particles in the upper levels).  The single-particle wave
functions come from a harmonic oscillator with length parameter $b=2.12$ fm.

The Hamiltonian, constructed from the generators of $SO(5)\times SO(5)$ (one
$SO(5)$ for each degenerate set of levels), is \cite{dussel70}:
\vspace{.2in}
\begin{eqnarray}
\label{eq:H}
H=\epsilon
\hat{N}_2~~~~~~~~~~~~~~~~~~~~~~~~~~~~~~~~~~~~~~~~~~~~~~~~~~~~~~~~~~~~~&&  \\
-G   \sum_{a,b=1}^2 \left( S^{\dag a}_{pp} S^b_{pp} +
S^{\dag a}_{nn} S^b_{nn} + g_{pp} S^{\dag a}_{pn} S^b_{pn} 
  - g_{ph}
\svec{T}_a \cdot \svec{T}_b \right) ~,&& \nonumber
\end{eqnarray}
where $a,b=1,2$ label the shells (lower and upper), $\epsilon$ is the energy
difference between the shells, $\hat{N}_2$ is the number operator for the
upper
shell, $\svec{T}_a$ is total isospin operator for shell $a$, and
\begin{eqnarray}
\label{eq:S}
S^{\dag a}_{pp} &= &\frac{1}{2}\sum_{\alpha \in a} \hat{j}_{\alpha}
[\pi^{\dag}_{\alpha} \pi^{\dag}_{\alpha}]^0_0 \nonumber \\
S^{\dag a}_{nn} &= &\frac{1}{2}\sum_{\alpha \in a} \hat{j}_{\alpha}
[\nu^{\dag}_{\alpha} \nu^{\dag}_{\alpha}]^0_0 \nonumber \\
S^{\dag a}_{pn} &= &\frac{1}{\sqrt{2}}\sum_{\alpha \in a}
\hat{j}_{\alpha}[\pi^{\dag}_{\alpha} \nu^{\dag}_{\alpha}]^0_0  ~.
\end{eqnarray}
Here $\pi^{\dag}_{\alpha}$ ($\nu^{\dag}_{\alpha}$) creates a proton
(neutron) in level $\alpha$ with angular momentum $j_{\alpha}$, $\hat{j}
\equiv \sqrt{2j+1}$, and the square brackets indicate angular-momentum
coupling. The algebra of $SO(5)$ contains the three pair-creation operators
above (for a given set $a$ of levels), three corresponding destruction
operators, the three components of the isospin $\svec{T}_a$, and the number
operator $\hat{N}_a$. Since $H$ contains only generators of $SO(5) \times
SO(5)$,
its lowest lying eigenstates will consist of configurations in which the
nucleons are entirely bound in isovector $S$ pairs of the type in Eq.\
(\ref{eq:S}).

The Hamiltonian in Eq.\ (\ref{eq:H}) is not the most complicated that could be
built from generators of $SO(5) \times SO(5)$.  One could, for instance,
assign
different strengths to the pairing forces in the two levels, or make the
proton
pairing stronger than the neutron pairing.  Even without the last
complication,
the Hamiltonian  violates isospin ($T$) conservation unless $g_{pp}=1$.  In our
application of this model, we will make an analogy between double Fermi decay,
the $2\nu$ version of which vanishes when isospin is exactly
conserved, and double-Gamow-Teller decay in a more general model, which
vanishes when Wigner $SU(4)$ is exactly conserved.  Of course, $SU(4)$ is more
badly violated than isospin, which is nearly a good quantum number, but in our
simplified model, we will let isospin be badly violated to mock up the
violation of $SU(4)$ in more realistic calculations\footnote{$SU(4)$ is
violated primarily at the mean-field level, while in our calculations isospin is
violated only in the two-body interaction.}.  When $T$ is violated, the space
of fully-paired states in the two sets of levels $pfg_{9/2}$ and $sdg_{7/2}$
has dimension 1042 for $^{76}$Ge and 1347 for the daughter nucleus $^{76}$Se.

We will calculate the ground-state to ground-state transition matrix elements
of two operators:  the $2\nu$ double-Fermi operator and the $0\nu$
double-Fermi operator (both obtained from the closure approximation).  These
operators have the form
\begin{equation}
\label{eq:ops}
\mathcal{M}_{\kappa} = \sum_{i,j} \mathcal{O}_{\kappa}(i,j) \tau^+_i \tau^+_j~,
\end{equation}
where  $i,j$ refer to particles and $\tau^+$ changes a neutron into
a proton.  The space/spin parts of the operators are
\begin{equation}
\label{eq:op2}
\mathcal{O}_{2\nu}(i,j) = 1~,
\end{equation}
and
\begin{equation}
\label{eq:op0}
\mathcal{O}_{0\nu}(i,j) = \frac{1}{|\svec{r}_i-\svec{r}_j|}~.
\end{equation}
The label $\kappa$ in Eq.\ (\ref{eq:ops}) refers to the kind of decay. 
The denominator in Eq.\ (\ref{eq:op0}) reflects the propagation of a 
virtual neutrino inside the nucleus.  We  have neglected the average nuclear 
excitation and short-range correlations in writing $\mathcal{O}_{0\nu}$.   We
will also 
evaluate an operator that includes those effects:
\begin{equation}
\label{eq:op0'}
\mathcal{O}^{\prime}_{0\nu}(i,j) = \frac{f(|\svec{r}_i-\svec{r}_j|)}
{|\svec{r}_i-\svec{r}_j|}~,
\end{equation}
where \cite{engel88,miller76}
\begin{equation}
\label{eq:src}
f(r) \equiv e^{-1.5 \bar{E}r/\hbar c} [1-e^{-\gamma_1r^2}(1-\gamma_2r^2)]^2~,
\end{equation}
with $\gamma_1=1.1$ fm$^{-2}$, $\gamma_2=0.68$ fm$^{-2}$, and $\bar{E}$ taken
somewhat arbitrarily to be 15 MeV.

The fully-paired basis states can be labelled
$|\mathcal{N}_1,T_1,M_1;\mathcal{
N}_2,T_2,M_2\rangle$, where $\mathcal{N}_1$ refers to the number of pairs in
level-set 1 (the $pfg_{9/2}$ shell), $T_1$ to the isospin of those
$\mathcal{N}_1$ pairs (with $\mathcal{N}_1-T_1$ even), $M_1$ to the
isospin projection of those pairs, etc., and $M_1+M_2 = 1/2(Z-N)$.  The
Hamiltonian,
Eq.\ (\ref{eq:H}), and $2\nu$ decay operator, Eq.\ (\ref{eq:op2}),
are products of generators and one can evaluate their matrix elements, which
depend only on the quantum numbers of the pairs, following, eg., Ref.\
\cite{hecht67}.  But the $0\nu$ operators, Eqs.\ (\ref{eq:op0}) and
(\ref{eq:op0'}), contain position-dependent factors that do not belong to
the algebra.  Their matrix elements will depend on the wave functions of
the single-particle states that make up our basis and require more effort
to evaluate.  We have computed them by applying the generalized
Wigner-Eckart theorem \cite{hecht67}, which for matrix elements of an operator
$\mathcal{M}$ between fully paired states in $SO(5)$ is:
\begin{eqnarray}
\label{eq:WE}
\langle \Omega_a,\mathcal{N}_a, T_a, M_a |
\mathcal{M}^{(\omega_1,\omega_2)}_{\mathcal{
N}_0,T_0,M_0} |\Omega_a,\mathcal{N}^{\prime}_a, T^{\prime}_a,M^{\prime}_a
\rangle = && \nonumber \\
\times \langle (\Omega_a,0)||\mathcal{M}^{(\omega_1,\omega_2)}||(\Omega_a,0)
\rangle ~
\langle T^{\prime}_a M^{\prime}_a;T_0 M_0 | T_a M_a \rangle  && \nonumber \\
\times \langle (\Omega_a,0) \mathcal{N}^{\prime}_a T^{\prime}_a;
(\omega_1,\omega_2) \mathcal{N}_0 T_0 || (\Omega_a,0) \mathcal{N}_a T_a\rangle
&&~.
\end{eqnarray}
Here the extra ``quantum number" $\Omega_a$, omitted in the labeling scheme
discussed in the previous paragraph, is the half the total degeneracy of the
level-set $a$,
\begin{equation}
\label{eq:Om}
\Omega_a = \frac{1}{2} \sum_{\alpha \in a} \hat{j}^2_{\alpha}~,~~~a=1,2~,
\end{equation}
and labels the representations $(\omega_1=\Omega_a,\omega_2=0)$ of $SO(5)$ in 
which all
particles are fully paired.  More general representations, such as that
characterizing the operator in Eq.\ (\ref{eq:WE}), require two
nonzero labels $\omega_1$ and $\omega_2$.  The first factor on the right hand
side of that
equation is a reduced matrix element that depends only on $\Omega_a$ and the
operator quantum numbers $\omega_1$ and $\omega_2$.  All the dependence on
the initial and final number of particles in the system, the initial and
final isospin and isospin projection, and the isospin and
particle-number quantum numbers of the operator appear in the double-barred
$SO(5)$ ``reduced Clebsch-Gordan" coefficient (which is independent of the
isospin-projection quantum numbers) and an ordinary isospin Clebsch-Gordan
coefficient.

The work of of Ref.\ \cite{hecht67} allows us to decompose the operators
$\mathcal{M}_{0\nu}$ and $\mathcal{M}^{\prime}_{0\nu}$ into operators with good 
$SO(5) \times SO(5)$ quantum numbers, lists some of the $SO(5)$ Clebsch-Gordan
coefficients we need, and contains enough information to allow us to
construct (laboriously) those it doesn't list:  the coefficients for
isospin-changing
components of operators belonging to the $SO(5)$ representations $(2,0)$
and $(2,2)$.  Here we show expressions for matrix elements of 
double-charge-changing
two-body operators in the ``small
space" --- all particles in the degenerate $pfg_{9/2}$ levels --- the states of
which we write in the form  $|\mathcal{N},T,M\rangle$, where all quantum
numbers refer to the first set of levels ($a=1$) because the second set is 
empty.
We focus on these matrix elements because they are the most complicated and 
because they will determine the behavior of the effective decay
operators when we truncate the wave function to the small space (other matrix 
elements needed for the full calculation are in the appendix).  Defining 
summed
particle-particle and particle-hole-like matrix elements of an arbitrary
two-body operator $\mathcal{M}_{\kappa}$ that changes charge by two units
(acting, for
generality of notation, in either of the two sets of levels) as
\begin{eqnarray}
\label{eq:pp,ph} \mathcal{F}^{pp}_{\kappa}[a,b] &\equiv& \sum_{\alpha \in a,
\beta \in b} \hat{j}_{\alpha}
\hat{j}_{\beta} ~ \langle [\alpha \alpha]^0|\mathcal{O}_{\kappa}
| [\beta \beta]^0\rangle
\nonumber \\
\mathcal{F}^{ph}_{\kappa}[a,b] &\equiv& \sum_{\alpha \in a,\beta \in b} {\sum_J}
\hat{J}^2 ~ \langle [\alpha \beta]^J |\mathcal{O}_{\kappa}
 |[\alpha \beta]^J\rangle~,
\end{eqnarray}
where, as before, $a$ and $b$ refer to a degenerate sets of levels and
the
matrix elements are antisymmetrized, we have when all the particles are in the
lowest set
\begin{widetext}
\begin{eqnarray}
\label{eq:d2} \langle\mathcal{N},T \pm
2,M+2|\mathcal{M}_{\kappa}|\mathcal{N},T,M
\rangle =
\frac{\sqrt{T_> (T_> -1)(2\Omega-\mathcal{N}+T_> + 1)(2\Omega -\mathcal{N} -
T_<)(\mathcal{N}+T_> + 1)(\mathcal{N}-T_< ) }}
{8(2T_>-1) \Omega (\Omega -1)(2 \Omega +1)}~~~~~~~~~~~~~~&& \\
\times \langle T \pm 2~M+2;1~-1|T_>-1~M+1\rangle \langle T ~M;1~1|T_>
-1~M+1\rangle
\left( 2\mathcal{F}^{ph}_{\kappa}[1,1]-(2 \Omega-1)
\mathcal{F}^{pp}_{\kappa}[1,1] \right) ~,&& \nonumber
\end{eqnarray}
and
\begin{eqnarray}
\label{eq:d0}
\langle\mathcal{N},T,M+2|\mathcal{M}_{\kappa}|\mathcal{N},T,M \rangle = &
{\displaystyle\frac{\sqrt{T(T+1)}}{4\sqrt{6(2T-1)(2T+3)}\Omega (\Omega -1)(2
\Omega +1)}}
 \langle T~M;2~2|T~M+2\rangle~~~~~~~~~~~~~~~~~~~~~~~~~~~~ & \nonumber\\
& \times \Big\{ \left[ \mathcal{N}(\mathcal{N}-2\Omega)
-3(2\Omega+1)+(4\Omega+1)T(T+1)\right] ~2\mathcal{F}^{ph}_{\kappa}[1,1]
~~~~~~~~~~~~~~~&
\\
&~~~~~- \left[ (2\Omega-1)\mathcal{N} (\mathcal{
N}-2\Omega)-3\Omega(2\Omega+1)+(2\Omega+3)T((T+1) \right]~
\mathcal{F}^{pp}_{\kappa}[1,1]
\Big\}~.& \nonumber
\end{eqnarray}
\end{widetext}
Here $\Omega \equiv \Omega_1$, and $T_>$ ($T_<$) is the largest (smallest) of 
the isospins in the bra
and ket of Eq.\ (\ref{eq:d0}).
The eigenstates of the Hamiltonian Eq.\ (\ref{eq:H}) have good particle
number $\mathcal{N}$ and isospin projection $M$, but are mixtures of states
with different isospins $T$.  The same is true of the projection of the
wave functions onto the small space, so both the equations above are
important.  

The most important fact about Eqs.\ (\ref{eq:d2}) and (\ref{eq:d0}) is that
they depend on the two-body matrix elements of the operator through only two
quantities, $\mathcal{F}^{pp}[1,1]$ and $\mathcal{F}^{ph}[1,1]$, the
particle-particle
and
particle-hole (with some factors of $\hat{j}$ removed) matrix elements summed
over all orbits in the lower shell.  The action of any two-body
operator that changes charge by two units on fully paired states in that shell
can therefore be
constructed by
specifying just two parameters.  This simplicity reflects the nature of the 
underlying correlations.  The
interaction cannot create particle-hole excitations involving
different orbits.  
\section{\label{s:3}Matrix elements, truncation of wave functions, and
effective operators}

\begin{figure}[bht]
\includegraphics[width=3in]{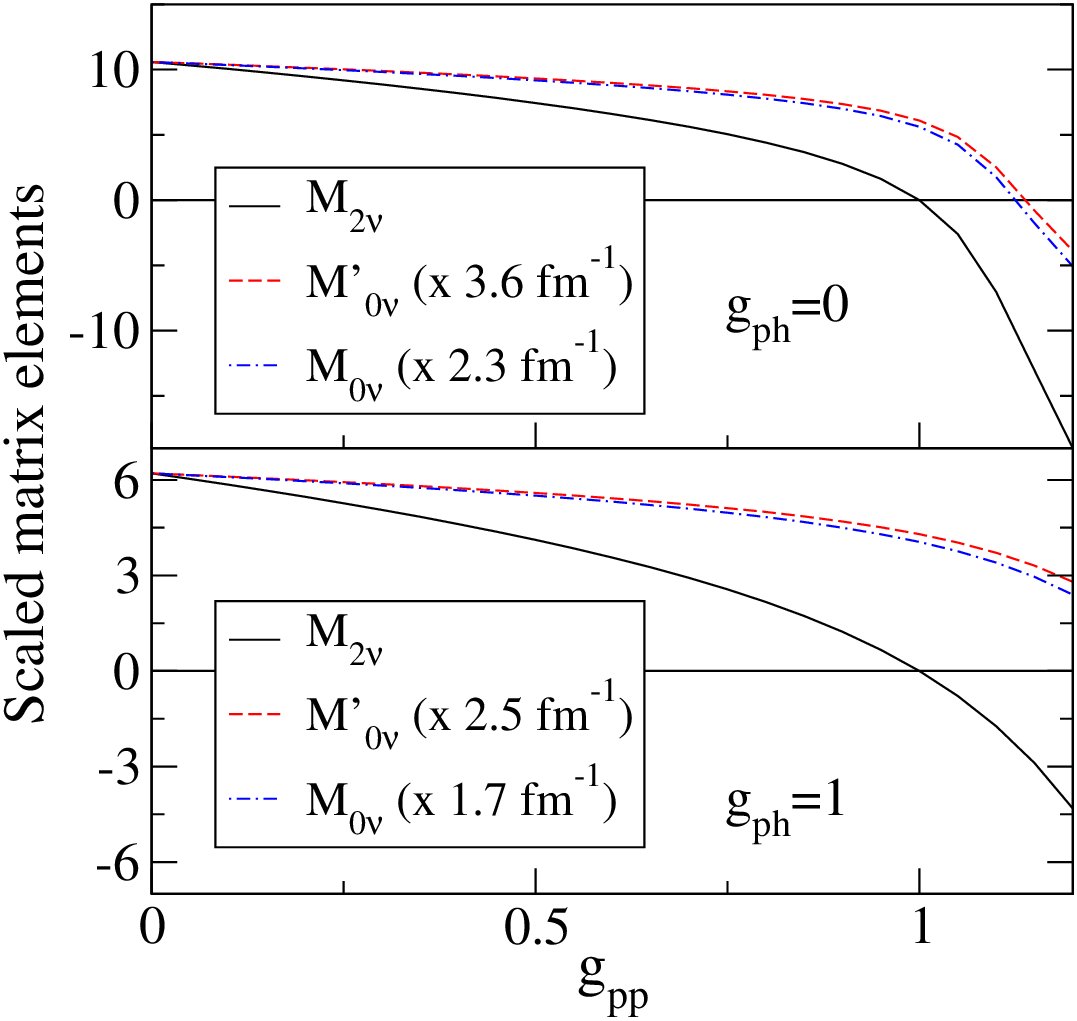}
\caption{\label{fig:1} Double-beta-decay matrix elements versus the strength
of neutron-proton pairing $g_{pp}$ for two values of the particle-hole
coupling $g_{ph}$.  The matrix elements of the two $0\nu$ operators are
scaled to that of the $2\nu$ operator at $g_{pp}=0$.}
\end{figure}

Before addressing the issues of effective decay operators in the small
space, we examine the behavior of the matrix elements 
\begin{equation}
\label{eq:medef}
M_{\kappa} \equiv \langle 0_f|\mathcal{M}_{\kappa} |0_i\rangle
\end{equation}
of the operators
$\mathcal{M}_{2\nu}$, $\mathcal{M}_{0\nu}$, and $\mathcal{M}^{\prime}_{0\nu}$
connecting the ground states of $^{76}$Ge and $^{76}$Se.  The matrix
elements are plotted in Fig.\ (\ref{fig:1}) for fixed $G$, $\epsilon=10G$
(resulting in about 20\% of the nucleons occupying the upper set of
levels), and two values of $g_{ph}$.  We vary $g_{pp}$, which
measures the strength of proton-neutron pairing (the graphs look
qualitatively the same when we change $\epsilon$ to $20G$ and about 6\%
of the particles are in the upper set.) To compare the behavior of the
three matrix elements with $g_{pp}$, we have scaled the $0\nu$ matrix
elements so that they are equal to the $2\nu$ elements at $g_{pp}=0$.  At
$g_{pp}=1$ isospin is conserved and the $2\nu$ matrix element crosses
zero.  In realistic calculations the analog of $g_{pp}$ in the
$J^{\pi}=1^+$ channel is close to 1; indeed, early QRPA calculations gave
matrix elements that were much too large precisely because they neglected
the particle-particle interaction.  The $0\nu$ matrix elements cross zero
only for larger $g_{pp}$, much larger when
$g_{ph}=1$, and are therefore
less sensitive than the $2\nu$ elements to that parameter when it is at 
a realistic value.
This reduced sensitivity is another aspect of our model that mirrors
more complicated calculations.   The two versions of the $0\nu$ matrix element 
are
nearly proportional; the short-range correlations and average excitation
energy shrink the matrix element by a factor of about 1.5, almost
independently of $g_{pp}$.
\vspace{.35in}

\begin{figure}[hbtp]
\includegraphics[width=3in]{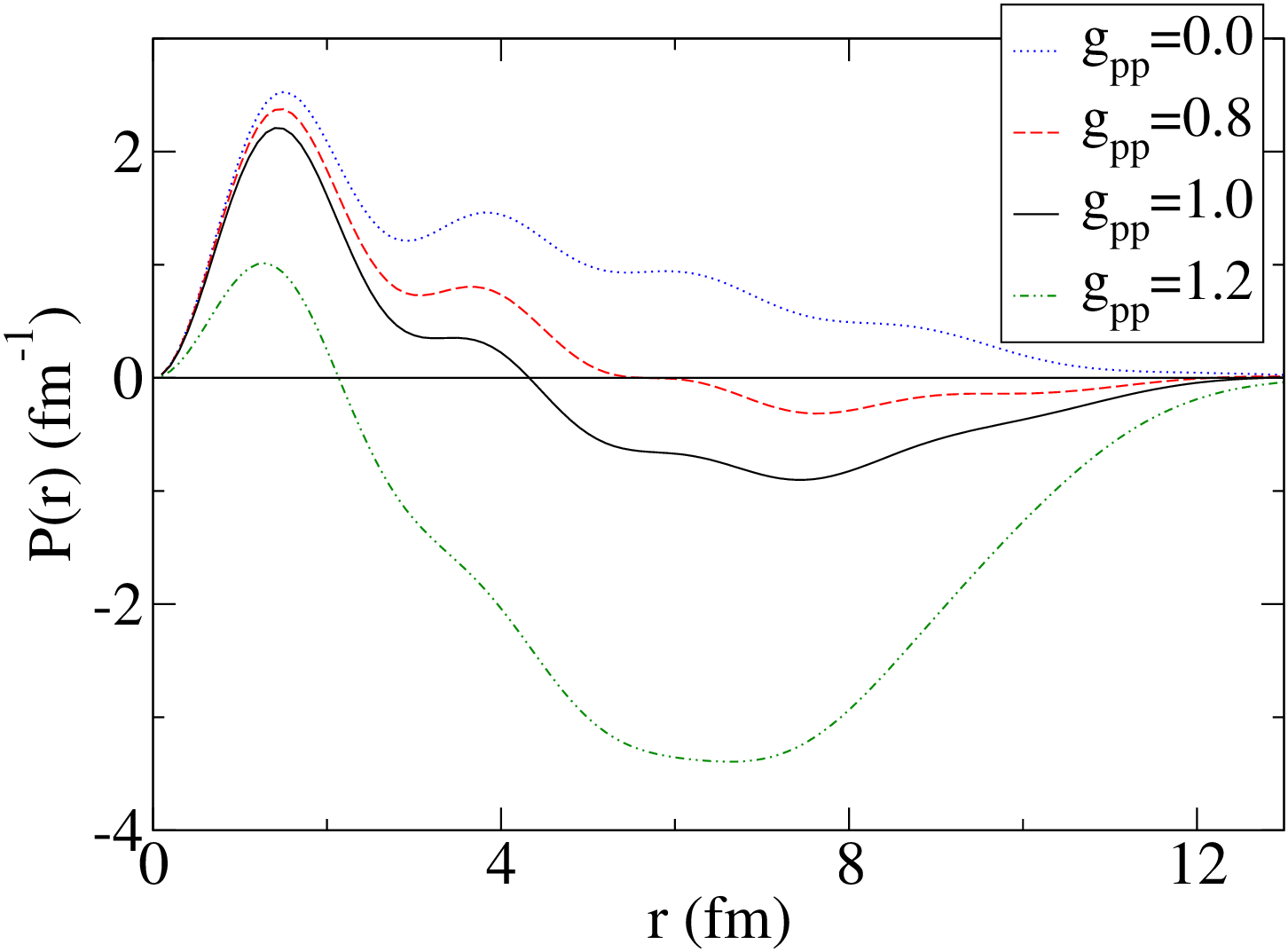}
\caption{\label{fig:2} The transition density $P(r)$ (see eqn.\
(\ref{eq:P})) as a function of internucleon separation $r$ for several
values of $g_{pp}$, with $\epsilon = 10 G$ and $g_{ph}=0$.}
\end{figure}

The different behavior of the $2\nu$ and $0\nu$ matrix elements with
$g_{pp}$ is explored in Fig.\ (\ref{fig:2}),
which shows the pair-separation transition density
\begin{equation}
\label{eq:P}
P(r) \equiv \langle 0_f | \sum_{i,j} 
\delta\left(r-|\svec{r}_i-\svec{r}_j|\right) \tau^+_i \tau^+_j | 0_i \rangle ~,
\end{equation}
($r$ is the magnitude of the internucleon separation) at several values of
$g_{pp}$.  The density $P(r)$ is
defined so that
\begin{eqnarray}
\int P(r) dr & =M_{2\nu} \\
\int \frac{P(r)}{r} dr & = M_{0\nu} ~,
\end{eqnarray}
etc.  The figure shows that as $g_{pp}$ increases $P(r)$ decreases and then
changes sign at large $r$, while remaining positive at small $r$.  At
$g_{pp}=1$ the integral of $P(r)$ is zero but the $1/r$-weighted integral
is still positive because small values of $r$ are most important.
Thus, the matrix elements of $\mathcal{M}_{0\nu}$ and
$\mathcal{M}^{\prime}_{0\nu}$ cross zero at larger $g_{pp}$ than that of
$\mathcal{M}_{2\nu}$. 

The reason
for that $P(r)$ is most affected by $g_{pp}$ at large $r$ is the following:  At
$g_{pp} = 0$ only one pair can participate
in the ground-state to ground-state decay; since there are no correlated
proton-neutron pairs, the decay operator turns two neutrons within a given pair
into two protons.  As $g_{pp}$ becomes larger the ground states contain
neutron-proton pairs, and double-beta decay can proceed, e.g, through the
transformation of one neutron
in a neutron-neutron pair and one in a (separate) proton-neutron pair.  These
two neutrons are farther apart on average than two in the same correlated pair,
so the transition density changes more at large $r$ than at small $r$, where it
stays positive.  
\vspace{.25in}

We now turn to the question of effective operators in the small space (all
nucleons in the $pfg_{9/2}$ shell).  The Bloch-Horowitz equation
\cite{bloch58},
\begin{equation}
H_\mathrm{eff}(E) = P H P + P H Q \frac{1}{E-QH}Q H P \quad,
\end{equation}
and approximations thereto are designed to construct an effective
Hamiltonian $H_\mathrm{eff}$ that obeys
\begin{equation}
\label{eq:Heff} H_\mathrm{eff}(E_k) P |\Psi_k\rangle=  P H | \Psi_k\rangle =
E_k P |\Psi_k\rangle~,
\end{equation}
where $H$ is the full Hamiltonian, $|\Psi_k\rangle$ is an eigenstate with
energy $E_k$, $P$ projects onto the subspace in which all particles are in the
$pfg_{9/2}$ shell, and $Q=1-P$.  Once the
projections of the eigenstates are known, effective transition operators
$\mathcal{M}^\mathrm{eff}_{\kappa}$ can be constructed to satisfy
\begin{equation}
\label{eq:Meff}
\langle \overline{\Psi}_k | \mathcal{M}^\mathrm{eff}_{\kappa}
|\overline{\Psi}_l\rangle = \langle \Psi_k |\mathcal{M}_{\kappa}
|\Psi_l\rangle~,
\end{equation}
where
\begin{equation}
\label{eq:barwf}
|\overline{\Psi}\rangle \equiv \frac{P| \Psi\rangle}{\sqrt{\langle
\Psi|P|\Psi\rangle}}~
\end{equation}
is the normalized component of $|\Psi\rangle$ that has all the particles in the
lower set of levels.  We don't actually have to know $H_\mathrm{eff}$ to
construct
$\mathcal{M}^\mathrm{
eff}_{\kappa}$; we just need the full wave functions $|\Psi_k\rangle$ and their
normalized projections $|\overline{\Psi}_k\rangle$.

Effective operators must in general be sums of 1-body, 2-body, \ldots up to
N-body terms.  It is common practice, however, to attempt to incorporate the
effects of the excluded space in an effective operator of the same rank
(number of bodies) as the original operator.  Thus effective electromagnetic
transition operators are often taken to be one-body but of a more general form
than the original operator, with ``level-dependent effective charges". 
In calculations of $2\nu$ double-beta decay, people often just multiply the bare
operator by a single effective coupling, usually $1/(1.25^2)$, to take into 
account
the observed quenching of spin-dependent transitions.  When calculating $0\nu$
decay they either use the bare operator or the same quenched coupling as for
$2\nu$ decay.  Nothing beyond a simple multiplicative renormalization has ever
been attempted.

\begin{figure}[htp]
\includegraphics[width=3in]{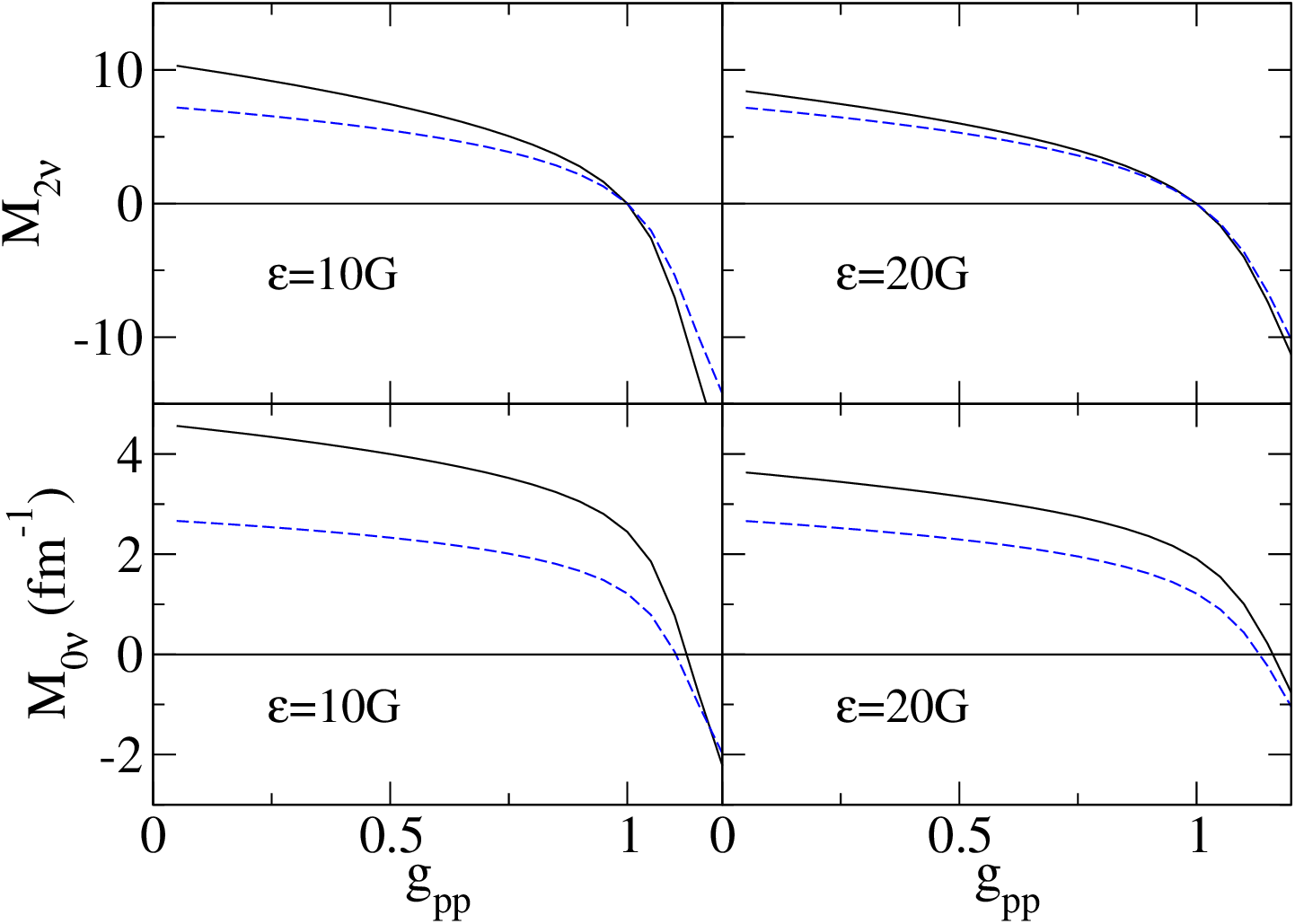}
\caption{\label{fig:3} $2\nu$ (top) and $0\nu$ (bottom) matrix elements
versus the strength of neutron-proton pairing $g_{pp}$ for $g_{ph}=0$.  The
solid lines are the results in the full models space and the dashed lines
are the results obtained by sandwiching the bare operator between normalized
projections of the wave functions onto the small
model space.}
\end{figure}

Although by definition there always exists a multiplicative factor that
will work, the trick is to know what it is.  Figure \ref{fig:3} shows that the
appropriate factor is different in $0\nu$ decay than
in $2\nu$ decay.  At
$g_{pp}=0.9$, a value yielding a reasonable approximation to the real world
within our model,  the renormalization factor for $0\nu$
decay must be 1.8 when $\epsilon=10 G$ and 1.5 when $\epsilon=20 G$
($g_{ph}=0$). 
For $2\nu$ decay, these numbers are 1.3 and 1.1 (adding the particle hole force
changes the
numbers only a little).  These results measure the quality
both of calculations that use bare operators and of those that attempt to
incorporate the effects of neglected single-particle levels by renormalizing
the
$2\nu$ and $0\nu$ operators by the same factor.  

\begin{figure}[bht]
\includegraphics[width=3in]{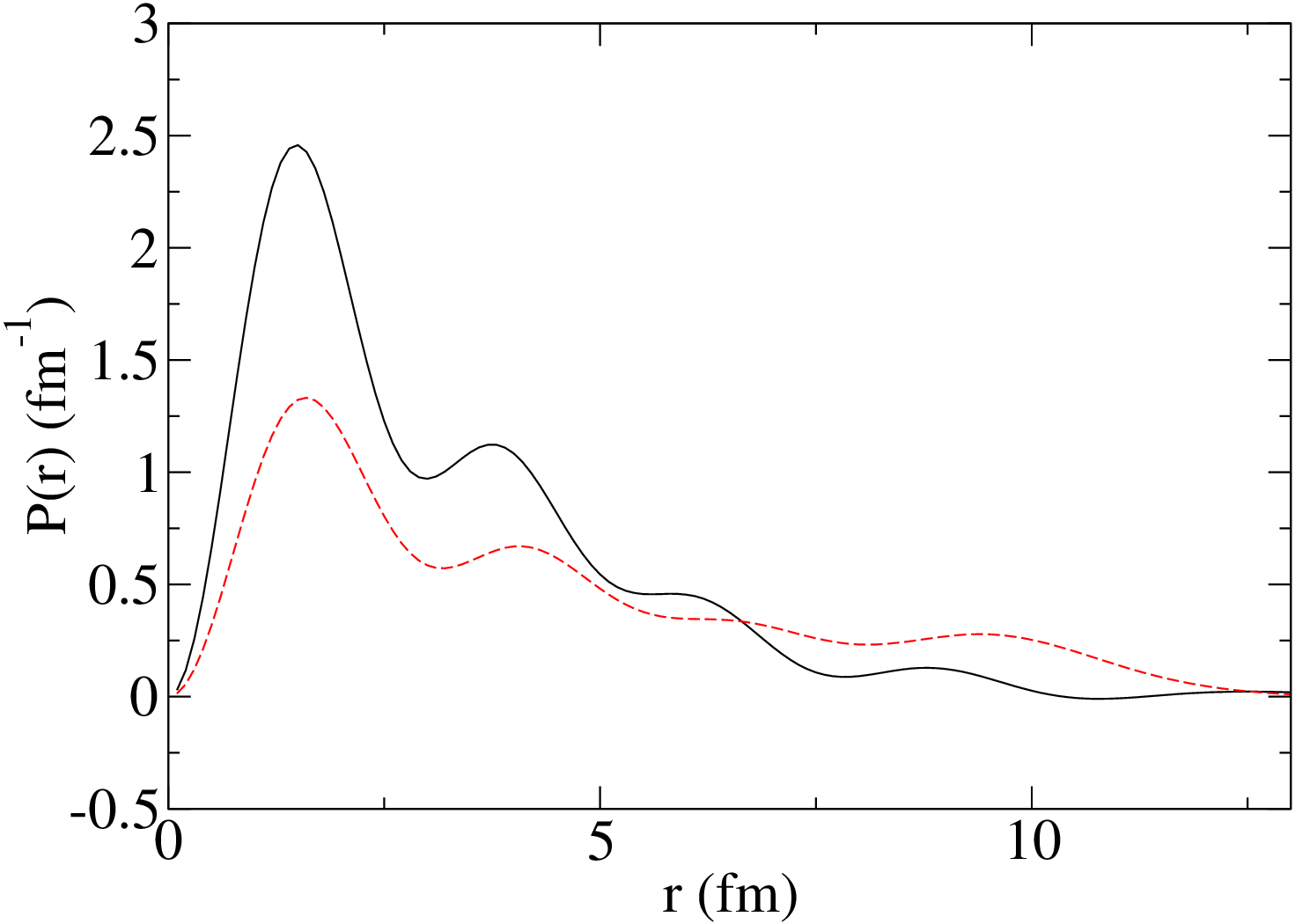}
\caption{\label{fig:4} The transition density $P(r)$ as a function of 
internucleon separation, at $g_{pp}=0.5$, for the full ground-state wave
functions 
(solid line) and their normalized projections onto the small space (dashed 
line).}
\end{figure}

Figure \ref{fig:4}, which shows $P(r)$ at $g_{pp}=0.5$ for the full ground
states $|0\rangle$ and their normalized small-space projections  
$|\overline{0}\rangle$, sheds some light on the different renormalization 
factors.  Interestingly, the transition density spreads out when projected onto 
the small space, so that most of the change is at small $r$ and, as a result,
the renormalization factor for the $0\nu$ 
transition is larger than for the $2\nu$ transition.  One can understand the
spreading by 
looking at the structure of the correlated pairs in Eq.\ (\ref{eq:S}). 
The one-body piece of the mean-square internucleon distance $\langle 
(\svec{r}_1-\svec{r}_2) \cdot (\svec{r}_1-\svec{r}_2) \rangle$ is 
uniformly 
reduced by the cross term $-2 \langle \svec{r}_1 \cdot \svec{r}_2 \rangle$.  In 
the $pfg_{9/2}$ shell $\svec{r}$ connects only the $f_{7/2}$ and $g_{9/2}$
levels.  
Adding the upper shell introduces many more contributions to $\langle \svec{r}_1 
\cdot \svec{r}_2 \rangle$ and so reduces the mean-square radius below its value
in one shell.  Alternatively
(and more simply), the $0\nu$ matrix element is more sensitive to occupation of
the upper set of levels because the neutrino propagator in Eq.\ (\ref{eq:op0})
allows the $0\nu$ operator to connect those levels with the more highly
occupied levels in the lower shell.   The $2\nu$ operator, which lacks radial
dependence, does not connect levels from different shells.
\vspace{.05in}

If simple renormalization by a constant factor
is dangerous, what about applying ideas used in 
electromagnetic physics to obtain a
more general two-body operator?  The usual practice there is to assume $N$
nucleons in the valence shell when constructing an effective $N$-body
operator, and then to apply that operator when there are larger numbers of
valence particles.  We can implement that procedure easily and exactly
here,
with $N = 2$.  
The ground-state to ground-state transition for two valence
neutrons in the small space decaying to two protons doesn't depend on the
parameter
$\mathcal{F}^{ph}_{\kappa,\mathrm{eff}}$ of the effective operator in Eq.\
(\ref{eq:Meff}) (i.e.\ the coefficient of $\mathcal{F}^{pp}_{\kappa}[1,1]$ in
Eq.\ (\ref{eq:d0}) vanishes when $\mathcal{N}=1$ and $T=1$ because the
particle-hole piece of the operator doesn't act on nucleons within the
same pair), and so
determines the parameter $\mathcal{F}^{pp}_{\kappa,\mathrm{eff}}$.  The excited
states in 
both
the initial and final nuclei have seniority two and are unaffected by the
interaction; their structure plus that of the paired ground state determine
$ \mathcal{F}^{ph}_{\kappa,\mathrm{eff}}$ in a simple way.  More precisely, the
effective
particle-particle and particle-hole parameters for the operator
$\mathcal{M}^\mathrm{eff}_{\kappa}$ are given by
\begin{equation}
\label{eq:2a}
\mathcal{F}^{pp}_{\kappa,\mathrm{eff}}=\mathcal{F}^{pp}_{\kappa}[1,1]~
\frac{\langle 0_f | \mathcal{M}_{\kappa}
| 0_i \rangle}{
\langle \overline{0}_f | \mathcal{M}_{\kappa} | \overline{0}_i \rangle}~,
\end{equation}
and
\begin{equation}
\label{eq:2b}
\mathcal{F}^{ph}_{\kappa,\mathrm{eff}}=\mathcal{F}^{ph}_{\kappa}[1,1] +
\frac{1}{2\Omega} \left(\mathcal{F}^{pp}_{\kappa,\mathrm{eff}} -
\mathcal{F}^{pp}_{\kappa}[1,1] \right)~,
\end{equation}
where the wave functions $| 0_i \rangle$  ($| 0_f \rangle$) are the full
two-neutron (two-proton) ground-state wave functions, the barred wave
functions are their normalized projections onto the small space, and
$\Omega\equiv
\Omega_1$ characterizes the $pfg_{9/2}$ shell.   With these identifications (the
$\mathcal{F}[1,1]$'s are ``bare" coupling constants), the
operator $\mathcal{M}_{\kappa}^\mathrm{eff}$ reproduces 
all transitions involving the lowest
$\Omega (2\Omega-1)$ states in each nucleus.  It is what one would obtain
by carrying the usual diagrammatic perturbation theory for the
two-valence-nucleon system to all orders.

Of course we really ought to be determining the effective operator for
$^{76}$Ge $\longrightarrow^{76}$Se, rather than the two-valence-nucleon
transition $^{42}$Ca $\longrightarrow^{42}$Ti.  That, however, would amount
to solving the problem exactly.  But one might decide that it would be
better (and feasible) to focus on ground-state to ground-state transitions,
in the two-nucleon system and, say, in the four-nucleon system $^{44}$Ca
$\longrightarrow^{44}$Ti.  Reproducing these two transitions would
determine the two parameters $\mathcal{F}^{pp}_{\kappa,\mathrm{eff}}$ and 
$\mathcal{F}^{ph}_{\kappa,\mathrm{eff}}$, and
would have the advantage of incorporating neutron-proton pairing, which we
know plays an important role in the heavier systems when $g_{pp}$ is turned
on, into the effective operator (through the two-proton two-neutron state
in $^{44}$Ti).

\begin{figure}[t!]
\includegraphics[width=3in]{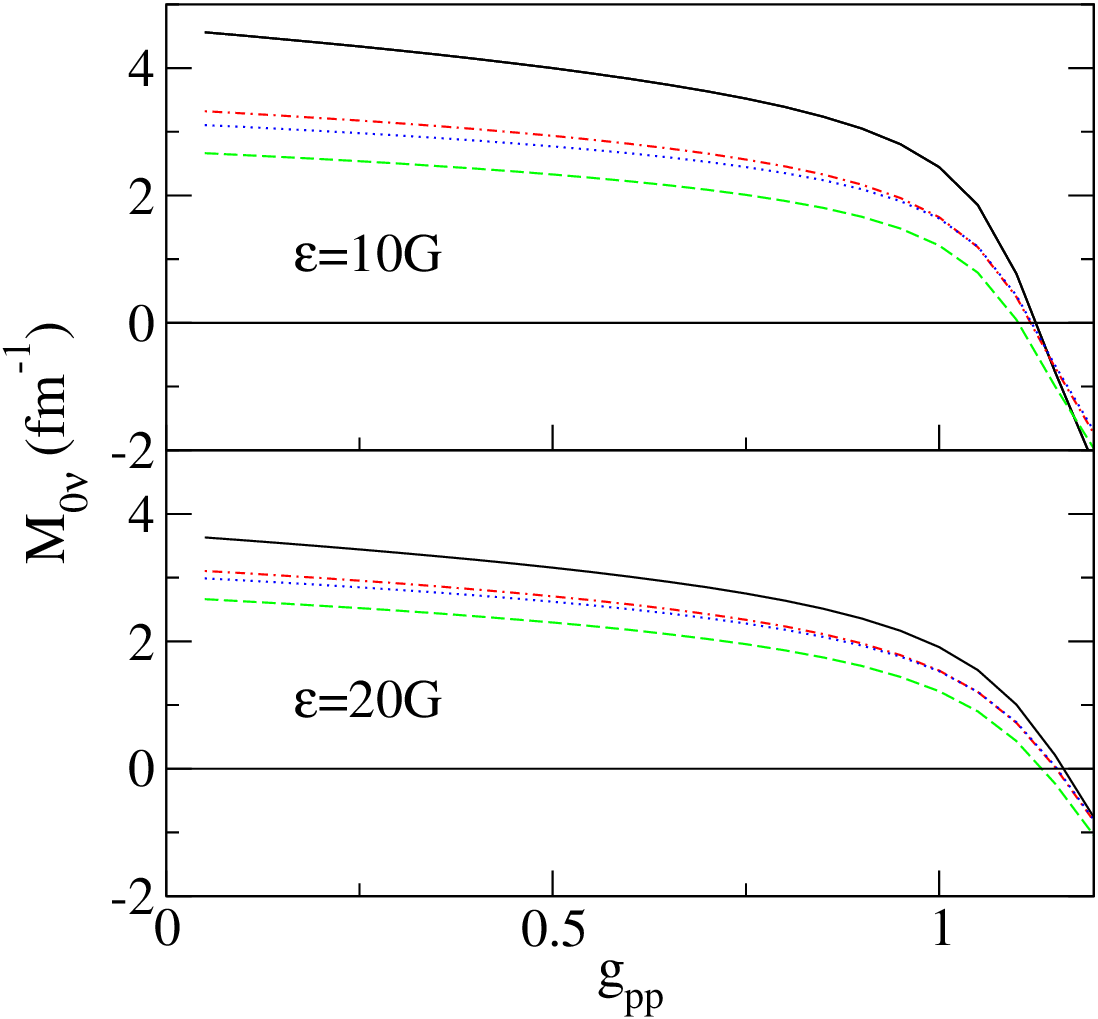}
\caption{\label{fig:5} Matrix elements of  $0\nu$ effective operators as a
function of $g_{pp}$.  The dashed line corresponds to the
bare operator in the small model space, the dotted line to the effective
operator that reproduces transitions in the two-valence-nucleon system, and
the dot-dashed line to the operator that reproduces ground-state transitions
in two- and four-valence-nucleon systems.  The solid line is the exact result
in the full model space.}
\end{figure}

We display the results of both these prescriptions for the operator
$M_{0\nu}$ in Fig.\ \ref{fig:5}.  Both procedures improve on the bare
operator, particularly when $\epsilon=20G$, where they make up about half
the difference between the bare and full results, but neither is
perfect\footnote{For $2\nu$ decay the bare operator gives a better
approximation than for $0\nu$ decay (see Fig.\ \ref{fig:3}), but the
effective-operator procedures improve on it only slightly.}.
As expected, the fit to transitions involving two and four nucleons works
better than the operator constructed entirely in the two-valence-nucleon
system, but not remarkably so.  We tried determining the effective operator
entirely from four-valence-nucleon systems, but the improvement was
minimal.  The difference between the exact and approximate results is a
measure of the amount higher-body effective operators contribute to the
transitions.

\begin{figure}[htp]
\includegraphics[width=3in]{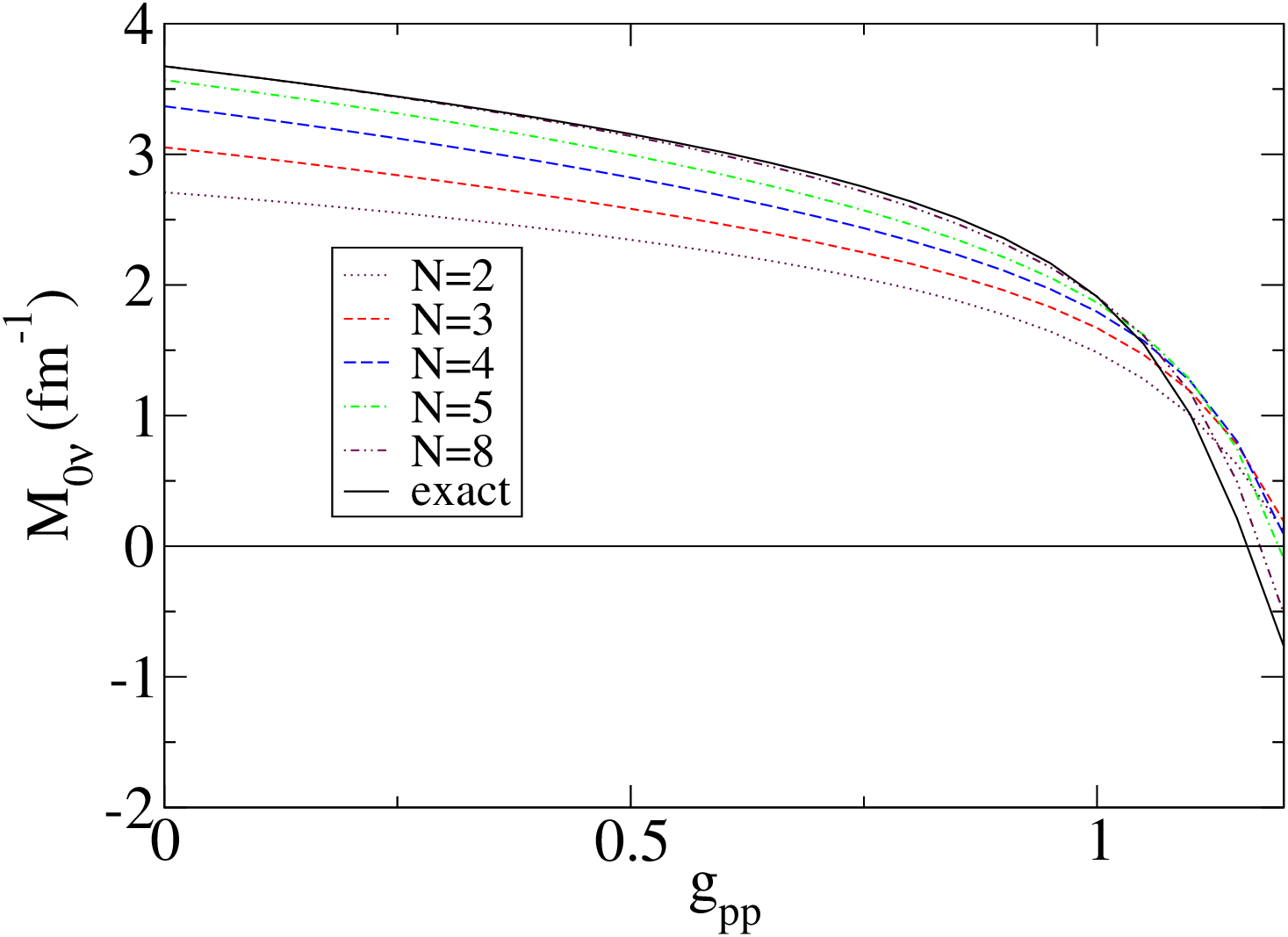}
\caption{\label{fig:6} Approximations to the $0\nu$ matrix element, for
$\epsilon =20G$, from a 
Lanczos representation of the Bloch Horowitz equation, as described in Ref.\
\cite{haxton00}.  Results are shown when the representation is truncated at
different numbers of 
Lanczos vectors, and compared to the exact results.}
\end{figure}

One reason that two-valence-nucleon systems are emphasized in the
traditional treatment of effective operators is that diagrammatic
perturbation theory becomes complicated when more particles are considered.
Reference \cite{haxton00} proposed a nonperturbative algorithm, based on the
Lanczos
representation of Green's functions as continued fractions \cite{haydock74},
that can be used for any number of particles. For
every state $|\mathcal{N},T,M \rangle$ in the small space and some guess
for the ground state energy $E$ (in both the initial and final nucleus), we
construct the approximation to the large space vector
$(E-QH)^{-1}|\mathcal{N},T,M \rangle$ and $H_\mathrm{eff}(E)$, and then
diagonalize the latter in the small space.  We then redefine $E$ to be the
lowest
eigen-energy and repeat the process until $E$ doesn't change.  Finally we
construct the Lanczos approximation to the full ground state wave functions
\begin{equation}
|0\rangle = \mathcal{Z} \left( 1+\frac{1}{E-QH} QH \right)|\overline{0}
\rangle \quad ,
\end{equation}
where $\mathcal{Z}$ is a computable normalization factor, and calculate the
transition matrix elements.  The results for different numbers of Lanczos
vectors appear in 
Fig.\ \ref{fig:6}, where $\epsilon = 20 G$.  Only 8 or 9 such vectors are 
required to get excellent results, but the iteration method fails to converge 
if $\epsilon$ is too small (e.g.\ $10 G$.).  Furthermore, it is hard in our 
context to see the advantage of this procedure over simply doing a Lanczos 
diagonalization to obtain the ground-state eigenvectors in the large space with 
the ``starting vector" determined, for example, by diagonalizing $H$ in the 
small space.  Such techniques are already exploited in shell-model calculations, 
and it's not clear that replacing them with the Bloch-Horowitz equation would be 
useful.  Reference \cite{haxton02} proposed a new perturbative approach to
solving that equation, but it has been worked out only in very light systems.
\vspace{.3in}

\section{\label{s:4} QRPA}
%

The QRPA is the most commonly used method for calculating double-beta decay
rates.  How accurate is it? If it is not so accurate, can its 
predictions
for $0\nu$ decay be improved by fixing certain parameters in the Hamiltonian to
reproduce $2\nu$ decay, as suggested in Ref.\ \cite{rodin03}?  We examine both
questions here.

We carry out the QRPA in the usual way (see, e.g., Ref.\ \cite{engel88}).  We
first solve the BCS equations in the initial and final nuclei, obtaining the
amplitudes
$u_{\alpha_p}^i,v_{\alpha_p}^i,u_{\alpha_n}^i,v_{\alpha_n}^i$
and  $u_{\alpha_p}^f,v_{\alpha_p}^f,u_{\alpha_n}^f,v_{\alpha_n}^f$. 
The corresponding occupation probabilities agree very well
with the exact occupation probabilities for all values of the
energy splitting $\epsilon$.
Next, we solve the usual QRPA equations of motion for both nuclei and use the
expressions in Ref.\ \cite{engel88} to evaluate
the transition matrix elements. 
(The only difference is that in that paper the matrix element was obtained by
averaging results from a calculation purely within the initial nucleus and from
one
within the final nucleus.   Here we use both 
sets
of $u$'s and $v$'s in the
same expression, inserting an overlap between intermediate states generated by
the QRPA from the initial and final ground states.) 
Since we evaluate only the closure matrix elements we can compare with the exact
solutions. The dependence
on the level splitting $\epsilon$ resides in the BCS amplitudes $u,v$
as well as in the QRPA amplitudes $X,Y$; the latter also
depend on the interaction strength $g_{pp}$.  The QRPA matrix elements are
always evaluated by expanding the transition operator in
proton-particle/neutron-hole multipole-mutipole form.  Since for $2\nu$ decay
the transition operator is just $2 \tau^+_1 \tau^+_2$, only the $0^+$
contribution
(in real nuclei, the $1^+$ contribution) exists.  The neutrino propagator causes
all multipoles contribute to $0\nu$ decay.

In our model the interaction
acts only in $0^+$ intermediate states. For all other multipoles $0\nu$
expressions reduce to their BCS form. 
The final result can be expressed as
\begin{eqnarray}  
\langle 0_f | \mathcal{M}_{\kappa} | 0_i \rangle_{\mathrm{QRPA}} = (0^+
~\mathrm{part})~~~~~~~~~~~~~~~~~~~~~ && \\ 
 + \sum_{\alpha,\beta,J} u_{\alpha_n}^f v_{\beta_p}^f u_{\beta_p}^i
v_{\alpha_n}^i   ([\beta\beta]^J|\mathcal{O}_{\kappa}|[\alpha\alpha]^J)
\nonumber \\ 
\times \left[
\hat{j}_{\alpha} \hat{j}_{\beta} \delta_{J,0}  -  \frac{\hat{J}^2
\delta_{\alpha,\beta}}{\hat{j}_{\alpha}^2}  \right] ~,  
~~~~~~~~~~~~~~~~~~&& \nonumber 
\label{eq:m0nu}
\end{eqnarray} 
where the matrix elements with rounded brackets are unsymmetrized (as usual in
the RPA) and ``$0^+$ part" refers to the matrix element when only the $0^+$
multipole in the particle-hole decomposition of the operator is included.  For
$2\nu$ decay the rest of the expresson vanishes, but for $0\nu$ decay there is
a simple contribution from the other multipoles that is independent of the
proton-neutron
interaction. 

\begin{figure}[htp]
\includegraphics[width=3in]{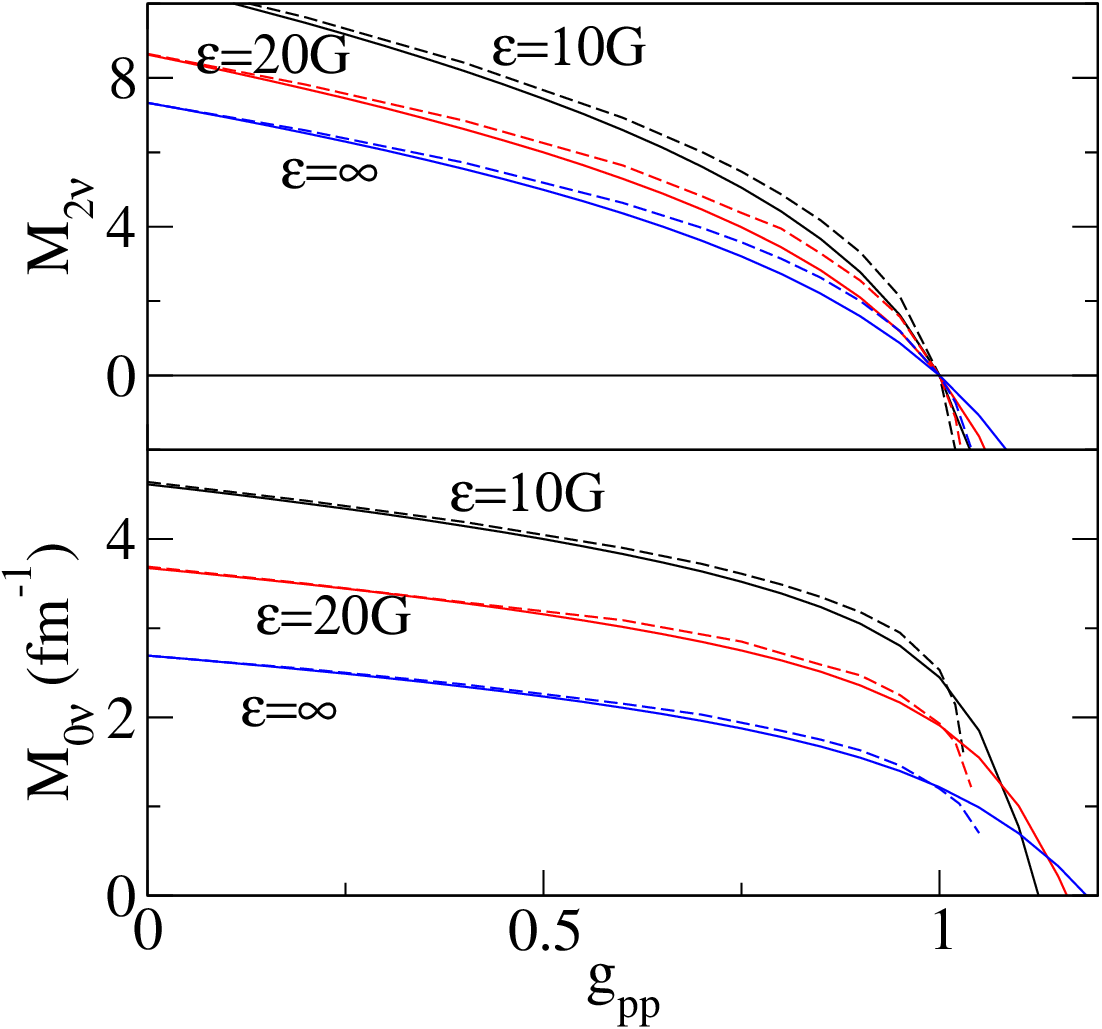}
\caption{Comparison of the exact and QRPA values
for the $2\nu$ matrix element (top) and $0\nu$ matrix element (bottom) for the
indicated values
of the spacing $\epsilon$ between the degenerate $fpg_{9/2}$ and
$sdg_{7/2}$ shells.
 }
\label{fig:7}
\end{figure}

The exact and QRPA results for both double-beta decay modes are compared in
Fig. \ref{fig:7}. The agreement is very good
over the whole interval  $g_{pp} = 0 - 1$. Note that
QRPA equations have no solution (the solutions are said to ``collapse") for
$g_{pp}$ only slightly
above unity.  The agreement is equally good for all values of the
splitting $\epsilon$. 
Previous comparisons between QRPA and exact results for $M_{2\nu}$ in this model
\cite{civitarese97} have not found such good agreement.   The better agreement
here
is related to the larger degeneracies $\Omega$ and the larger number of
particles we use to represent shell-model calculations in $^{76}$Ge.  

Reference \cite{rodin03} suggested a procedure for reducing the dependence of
the
$0\nu$ rate
on the number of single-particle states used in the QRPA. Simply put, the method
is
this:
assuming that the strength of the particle-particle interaction, $g_{pp}$,
is the most important parameter, one evaluates both matrix elements
$M_{0\nu}$ and $M_{2\nu}$ as a function of that parameter. One can then
invert the relation between $M_{2\nu}$ and $g_{pp}$ and
express $M_{0\nu}$ directly in terms of $M_{2\nu}$.  Reference
\cite{rodin03} did this within the
QRPA and in a modified version, the Renormalized QRPA \cite{toivanen95}, 
for $^{76}$Ge, $^{100}$Mo, $^{130}$Te, and $^{136}$Xe.
It turned out that the effects of single-particle levels far away from
the Fermi surface, while noticeable for $M_{0\nu}$ and $M_{2\nu}$ 
considered separately, became negligible when $M_{0\nu}$ was
plotted against  $M_{2\nu}$. In other words, knowing $M_{2\nu}$
(from experiment) one could obtain the unknown $M_{0\nu}$, and its value was the
same whether
the far away single-particle states were included or not in the QRPA, as long as
at least two oscillator shells (more than in our small space) were included.  
The question remains whether the
QRPA result, even if insensitive to the model space, is correct.

\begin{figure}[htp]
\includegraphics[width=3in]{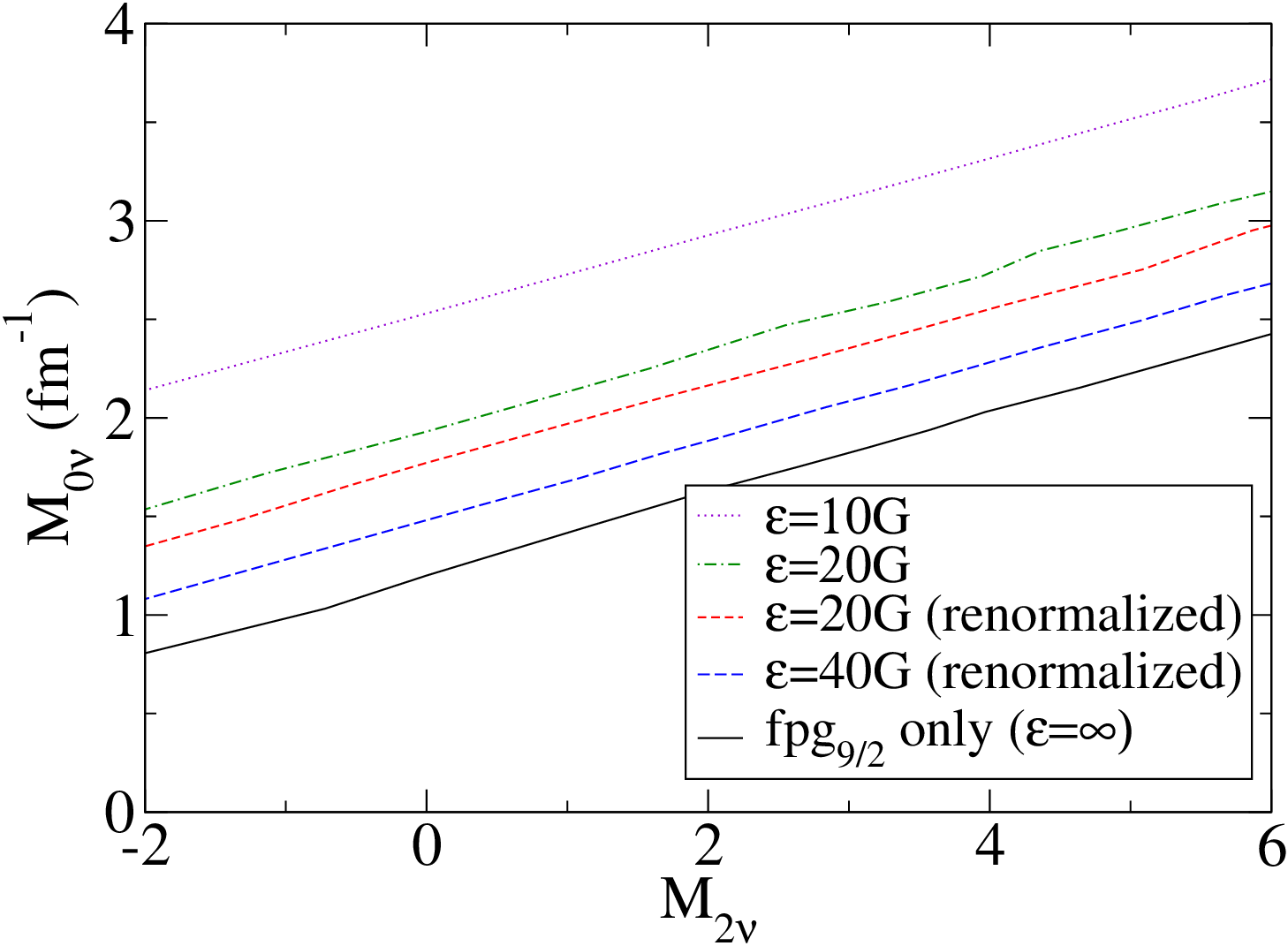}
\caption{Dependence of the matrix element $M_{0\nu}(g_{pp})$
on the matrix element  $M_{2\nu}(g_{pp})$ for the indicated values of
the level spacing $\epsilon$. See text for further explanation.}
\label{fig:8}
\end{figure}

When we carry out an analogous procedure in the $SO(5) \times SO(5)$ model,
we find that $M_{0\nu}$ and $M_{2\nu}$ are related
as shown in Fig.\ \ref{fig:8}. While the relation between
them is a simple one, it is not independent of $\epsilon$, as it would be if the
ideas of Ref.\ \cite{rodin03} played out perfectly.  Instead, the
offset in $M_{0\nu}$ depends on $\epsilon$ (but is
independent of $g_{pp}$). With increasing $\epsilon$ the lines in the figure,
for obvious reasons, come closer together.   While fixing $g_{pp}$ to the
exact (large-space) value of $M_{2\nu}$ improves the small-space prediction of
$M_{0\nu}$, it always leaves it smaller than the exact result.

In the realistic calculations the strength of the pairing interaction was
adjusted so that the pairing gap was the same in the ``large'' and
``small'' spaces.  We do the same at $\epsilon=20 G$; the corresponding line
moves closer to the $\epsilon=\infty$ limit, but is still significantly
different.  Finally, in Ref.\ \cite{rodin03} the proton-proton and
neutron-neutron pairing constants were adjusted separately. With a typical
value $G \sim 22/A {\rm ~MeV ~and~} \hbar \omega = 41/A^{1/3}$ MeV, the value
of $\epsilon$ corresponding to Ref.\ \cite{rodin03} is 
at least $40 G$. 
With that value and a charge-dependent pairing renormalization $G_{nn}/G_{pp}
\sim 1.2$
here, the line moves even closer to its limiting value.

The matrix elements in Fig.\ \ref{fig:8} were obtained without any
particle-hole
force, i.e.\ for $g_{ph} = 0.$ When the calculation is repeated with $g_{ph} =
1$ both $M_{0\nu}$ and $M_{2\nu}$ are significantly reduced (see Fig.\
\ref{fig:1}). When the
parameter $g_{pp}$ is eliminated, however, and $M_{0\nu}$ is expressed as a
function of $M_{2\nu}$,
the resulting lines are essentially identical to those shown in Fig.\
\ref{fig:8}. 
In other words,
the procedure of Ref.\ \cite{rodin03} eliminates the dependence of $M_{0\nu}$ on
the parameter $g_{ph}$,
at least in this model. This important conclusion should be tested in
more realistic QRPA calculations.

\vspace{.3in}

Why does the prescription of Ref.\ \cite{rodin03} fail to fully eliminate
model-space dependence in this model?
There are several possibilities.  As we pointed out earlier,
the magnitude of the level splitting $\epsilon$ used here is unrealistically
small. Using larger $\epsilon$ moves the large-space lines
closer to those of the small space. Also, the interaction here affects only the
$0^+$ multipole, while in realistic calculations all multipoles
are affected to some degree.
In any event, even with its shortcomings the procedure is interesting enough
here 
that we find
$M_{0\nu} = a \times M_{2\nu} + b$ independently of the values of both $g_{pp}$
and $g_{ph}$, with $b$ decreasing as the level splitting increases. Combined
with the accuracy of the QRPA in our model, this implies that the procedure
gives a lower limit for the exact value of
$M_{0\nu}$ provided that the Hamiltonian is corrected to
reproduce the exact value of $M_{2\nu}$.

%
\section{\label{s:5}Conclusions}
%
%
%

Our most important findings are that 1) calculated $2\nu$ and $0\nu$ decay rates
are affected differently by model-space truncation, 2) familiar procedures for
finding effective two-body decay operators to correct for truncation help
matters but leave substantial room for improvement, and 3) The procedure of
Ref.\ \cite{rodin03} for eliminating model-space dependence in the QRPA
helps but
doesn't work as well in our model as in realistic 
calculations.

Several issues still need to be investigated.  How accurate are the QRPA results
of Ref.\ \cite{rodin03} once the renormalization procedure is applied? 
Ironically, although the QRPA itself is very accurate in our model, 
almost certainly more accurate than in reality because the model has
very collective modes (the pairs), the renormalization procedure leaves more
residual model-space dependence here than in realistic calculations.  We
therefore can't really address the question in our model.   
Another issue we haven't addressed is the effects of
very-high momentum states on double-beta-decay operators.  Such effects are
usually
simulated by the short-range-correlation function $f$ in Eq.\ (\ref{eq:src}),
but it's not clear how accurate a result is achieved that way.  We are
currently examining two-body short-range correlations by summing
ladder diagrams similar to those that make up the Brueckner G-matrix.  

Properly
treating more-body effects will be harder, both for short-range correlations
and the longer-range correlations we investigated here.  Finding a good
approximation scheme to incorporate them into shell model calculations is, in
our view, the most important problem in the theoretical treatment of
double-beta decay.  Because only the ground states need to be calculated, much
of conventional effective-operator theory is superfluous and the biggest payoff
may come from pushing Lanczos-diagonalization methods to the largest model
spaces possible, leaving only high-momentum correlations to be absorbed into
operators.

\appendix
\section{}

Here we supplement Eqs.\ (\ref{eq:d2}) and (\ref{eq:d0}) with relations 
necessary to evaluate matrix elements of charge-changing two-body operators 
between states of the form  $|\mathcal{N}_1,T_1,M_1;\mathcal{
N}_2,T_2,M_2\rangle$, corresponding to particles in both sets of levels.  Eqs.\
(\ref{eq:d2}) and 
(\ref{eq:d0}) can be used with $\mathcal{N}=\mathcal{N}_1$ ($\mathcal{N}_2$), 
$T=T_1$ ($T_2$), $M=M_1$ ($M_2$) and the [1,1] ([2,2]) $\mathcal{F}$'s  whenever
$\mathcal{N}_2$ 
($\mathcal{N}_1$), $T_2$ ($T_1$), and $M_2$ ($M_1$) are the same in the bra as 
in the ket.  When those isospin quantum numbers change, we also need ({\it with} 
${M^{\prime}_1=M_1+1}$ {\it and} $M^{\prime}_2=M_2+1$ {\it
everywhere below}):
\begin{eqnarray}
\label{eq:a2} \langle \mathcal{N}_1,T^{\prime}_1,M^{\prime}_1; 
\mathcal{N}_2,T^{\prime}_2,M^{\prime}_2 | \mathcal{M}_{\kappa} |
\mathcal{N}_1,T_1,M_1; 
\mathcal{N}_2,T_2,M_2 \rangle = &&\nonumber \\
\frac{1}{2} \delta_{T_1,T^{\prime}_1} \delta_{T_2,T^{\prime}_2} \frac{\sqrt{T_1 
(T_1 +1) T_2 (T_2+1)}}{\Omega_1 \Omega_2} ~~~~~~~~~~~~~~~~~
&& \\
\times \langle T_1~M_1; 1~1 |T_1~M^{\prime}_1 \rangle \langle T_2~M_2; 1~1 
|T_2~M^{\prime}_2 \rangle ~ \mathcal{F}^{ph}_{\kappa}[12] ~,~~~~~&& \nonumber
\end{eqnarray}
where the delta function arises from the condition that $\mathcal{N}-T$ be even. 
When the operator moves pairs from one set of levels to another, finally, we 
need
\begin{eqnarray}
\label{eq:a3} \langle \mathcal{N}_1+1,T^{\prime}_1,M^{\prime}_1; 
\mathcal{N}_2-1,T^{\prime}_2,M^{\prime}_2 | \mathcal{M}_{\kappa} |
\mathcal{N}_1,T_1,M_1; 
\mathcal{N}_2,T_2,M_2 \rangle &&\nonumber \\
= ~ \frac{1}{8\Omega_1\Omega_2}  H^{\Omega_1}_{\mathcal{N}_1, T_1,T^{\prime}_1} 
H^{\Omega_2}_{\mathcal{N}_2-1, T^{\prime}_2,T_2} ~~~~~~~~~~~~~~~~~~~~~~~~~~~~~~
&& \\
\times \langle T_1~M_1; 1~1 |T^{\prime}_1~M^{\prime}_1 \rangle \langle 
T^{\prime}_2 ~ M^{\prime}_2; 1~-1 |T_2~M_2 \rangle ~
\mathcal{F}^{pp}_{\kappa}[12] 
~,~~~~~&& \nonumber
\end{eqnarray}
and
\begin{eqnarray}
\label{eq:a4} \langle \mathcal{N}_1-1,T^{\prime}_1,M^{\prime}_1; 
\mathcal{N}_2+1,T^{\prime}_2,M^{\prime}_2 | \mathcal{M}_{\kappa} |
\mathcal{N}_1,T_1,M_1; 
\mathcal{N}_2,T_2,M_2 \rangle &&\nonumber \\
= ~ \frac{1}{8\Omega_1\Omega_2}  H^{\Omega_1}_{\mathcal{N}_1-1, 
T^{\prime}_1,T_1} H^{\Omega_2}_{\mathcal{N}_2, T_2,T^{\prime}_2} 
~~~~~~~~~~~~~~~~~~~~~~~~~~~~~~
&& \\
\times \langle T^{\prime}_1~M^{\prime}_1; 1~-1 |T_1~M_1 \rangle \langle T_2 ~ 
M_2; 1~1 |T^{\prime}_2~M^{\prime}_2 \rangle ~ \mathcal{F}^{pp}_{\kappa}[12]
~,~~~~~&& 
\nonumber
\end{eqnarray}
where
\begin{equation}
\label{eq:a5}
H^\Omega_{\mathcal{N},T,T+1}= - \sqrt{\frac{(T+1) ( 2\Omega -\mathcal{N} -T) 
(\mathcal{N} +T +3)}{2T+3}}~,
\end{equation}
and
\begin{equation}
\label{eq:a6}
H^\Omega_{\mathcal{N},T,T-1}=  \sqrt{\frac{ T ( 2\Omega -\mathcal{N} +T +1) 
(\mathcal{N} -T +2)}{2T-1}}~.
\end{equation}

\begin{acknowledgments}
We thank J.N.\ Ginocchio and K.T.\ Hecht for helpful discussions about $SO(5)$. 
This work was supported in part by the U.S.\ Department of Energy under grants
DE-FG02-02ER41216, and DE-FG02-88ER-40397.
\end{acknowledgments}


\begin{thebibliography}{12}
\expandafter\ifx\csname natexlab\endcsname\relax\def\natexlab#1{#1}\fi
\expandafter\ifx\csname bibnamefont\endcsname\relax
  \def\bibnamefont#1{#1}\fi
\expandafter\ifx\csname bibfnamefont\endcsname\relax
  \def\bibfnamefont#1{#1}\fi
\expandafter\ifx\csname citenamefont\endcsname\relax
  \def\citenamefont#1{#1}\fi
\expandafter\ifx\csname url\endcsname\relax
  \def\url#1{\texttt{#1}}\fi
\expandafter\ifx\csname urlprefix\endcsname\relax\def\urlprefix{URL }\fi
\providecommand{\bibinfo}[2]{#2}
\providecommand{\eprint}[2][]{\url{#2}}

\bibitem[{\citenamefont{Elliot and Vogel}(2002)}]{elliot02}
\bibinfo{author}{\bibfnamefont{S.}~\bibnamefont{Elliot}} \bibnamefont{and}
  \bibinfo{author}{\bibfnamefont{P.}~\bibnamefont{Vogel}},
  \bibinfo{journal}{Ann. Rev. Nucl. Part. Sci.} \textbf{\bibinfo{volume}{52}},
  \bibinfo{pages}{115} (\bibinfo{year}{2002}).

\bibitem[{\citenamefont{Hirsch et~al.}(1997)\citenamefont{Hirsch, Hess, and
  Civitarese}}]{civitarese97}
\bibinfo{author}{\bibfnamefont{J.~G.} \bibnamefont{Hirsch}},
  \bibinfo{author}{\bibfnamefont{P.~O.} \bibnamefont{Hess}}, \bibnamefont{and}
  \bibinfo{author}{\bibfnamefont{O.}~\bibnamefont{Civitarese}},
  \bibinfo{journal}{Phys.\ Rev.\ C} \textbf{\bibinfo{volume}{56}},
  \bibinfo{pages}{199} (\bibinfo{year}{1997}).

\bibitem[{\citenamefont{Rodin et~al.}(2003)\citenamefont{Rodin, Faessler,
  \v{S}imkovic, and Vogel}}]{rodin03}
\bibinfo{author}{\bibfnamefont{V.~A.} \bibnamefont{Rodin}},
  \bibinfo{author}{\bibfnamefont{A.}~\bibnamefont{Faessler}},
  \bibinfo{author}{\bibfnamefont{F.}~\bibnamefont{\v{S}imkovic}},
  \bibnamefont{and} \bibinfo{author}{\bibfnamefont{P.}~\bibnamefont{Vogel}},
  \bibinfo{journal}{Phys.\ Rev.\ C} \textbf{\bibinfo{volume}{68}},
  \bibinfo{pages}{044302} (\bibinfo{year}{2003}).

\bibitem[{\citenamefont{Dussel et~al.}(1970)\citenamefont{Dussel, Maqueda, and
  Perazzo}}]{dussel70}
\bibinfo{author}{\bibfnamefont{G.~G.} \bibnamefont{Dussel}},
  \bibinfo{author}{\bibfnamefont{E.}~\bibnamefont{Maqueda}}, \bibnamefont{and}
  \bibinfo{author}{\bibfnamefont{R.~P.~J.} \bibnamefont{Perazzo}},
  \bibinfo{journal}{Nucl.\ Phys.} \textbf{\bibinfo{volume}{A153}},
  \bibinfo{pages}{469} (\bibinfo{year}{1970}).

\bibitem[{\citenamefont{Engel et~al.}(1988)\citenamefont{Engel, Vogel, and
  Zirnbauer}}]{engel88}
\bibinfo{author}{\bibfnamefont{J.}~\bibnamefont{Engel}},
  \bibinfo{author}{\bibfnamefont{P.}~\bibnamefont{Vogel}}, \bibnamefont{and}
  \bibinfo{author}{\bibfnamefont{M.~R.} \bibnamefont{Zirnbauer}},
  \bibinfo{journal}{Phys.\ Rev.\ C} \textbf{\bibinfo{volume}{37}},
  \bibinfo{pages}{731} (\bibinfo{year}{1988}).

\bibitem[{\citenamefont{Miller and Spencer}(1976)}]{miller76}
\bibinfo{author}{\bibfnamefont{G.~A.} \bibnamefont{Miller}} \bibnamefont{and}
  \bibinfo{author}{\bibfnamefont{J.~E.} \bibnamefont{Spencer}},
  \bibinfo{journal}{Ann.\ Phys.} \textbf{\bibinfo{volume}{100}},
  \bibinfo{pages}{562} (\bibinfo{year}{1976}).

\bibitem[{\citenamefont{Hecht}(1967)}]{hecht67}
\bibinfo{author}{\bibfnamefont{K.~T.} \bibnamefont{Hecht}},
  \bibinfo{journal}{Nucl.\ Phys.} \textbf{\bibinfo{volume}{A102}},
  \bibinfo{pages}{11} (\bibinfo{year}{1967}).

\bibitem[{\citenamefont{Bloch and Horowitz}(1958)}]{bloch58}
\bibinfo{author}{\bibfnamefont{C.}~\bibnamefont{Bloch}} \bibnamefont{and}
  \bibinfo{author}{\bibfnamefont{J.}~\bibnamefont{Horowitz}},
  \bibinfo{journal}{Nucl.\ Phys.} \textbf{\bibinfo{volume}{8}},
  \bibinfo{pages}{91} (\bibinfo{year}{1958}).

\bibitem[{\citenamefont{Haxton and Song}(2000)}]{haxton00}
\bibinfo{author}{\bibfnamefont{W.~C.} \bibnamefont{Haxton}} \bibnamefont{and}
  \bibinfo{author}{\bibfnamefont{C.~L.} \bibnamefont{Song}},
  \bibinfo{journal}{Phys.\ Rev.\ Lett.} \textbf{\bibinfo{volume}{84}},
  \bibinfo{pages}{5484} (\bibinfo{year}{2000}).

\bibitem[{\citenamefont{Haydock}(1974)}]{haydock74}
\bibinfo{author}{\bibfnamefont{R.}~\bibnamefont{Haydock}},
  \bibinfo{journal}{J.\ Phys.} \textbf{\bibinfo{volume}{A7}},
  \bibinfo{pages}{2120} (\bibinfo{year}{1974}).

\bibitem[{\citenamefont{Haxton and Luu}(2002)}]{haxton02}
\bibinfo{author}{\bibfnamefont{W.~C.} \bibnamefont{Haxton}} \bibnamefont{and}
  \bibinfo{author}{\bibfnamefont{T.}~\bibnamefont{Luu}},
  \bibinfo{journal}{Phys.\ Rev.\ Lett.} \textbf{\bibinfo{volume}{89}},
  \bibinfo{pages}{182503} (\bibinfo{year}{2002}).

\bibitem[{\citenamefont{Toivanen and Suhonen}(1995)}]{toivanen95}
\bibinfo{author}{\bibfnamefont{J.}~\bibnamefont{Toivanen}} \bibnamefont{and}
  \bibinfo{author}{\bibfnamefont{J.}~\bibnamefont{Suhonen}},
  \bibinfo{journal}{Phys.\ Rev.\ Lett.} \textbf{\bibinfo{volume}{75}},
  \bibinfo{pages}{410} (\bibinfo{year}{1995}).

\end{thebibliography}
\end{document}